\documentclass[12pt,a4paper]{article}
\usepackage{blindtext}
\usepackage[T1]{fontenc}
\usepackage[utf8]{inputenc}
\DeclareUnicodeCharacter{2212}{-}
\usepackage{amsmath}
\usepackage{graphicx}
\graphicspath{{images/}}
\usepackage{subfigure}
\usepackage{hyperref}
\usepackage{authblk}
\usepackage[normalem]{ulem}
\usepackage[stable]{footmisc}
\usepackage[superscript]{cite}
\usepackage{comment}
\hypersetup{
    colorlinks=true,
    linkcolor=blue,
    filecolor=magenta,      
    urlcolor=cyan,
}
\usepackage{vmargin}
\setmarginsrb{2.5 cm}{1.5 cm}{2.5 cm}{1.5 cm}{0.5 cm}{0.5 cm}{0.5 cm}{0.5 cm}

\date{\today}
\usepackage{setspace}
\usepackage{color,soul}

\title{\bf{Rouse Mode Analysis of Chain Relaxation in Reversibly Crosslinked Polymer Melts}}
\author[1,2]{Rahul Karmakar}

\affil[1]{Department of Chemical Engineering, Indian Institute of Technology Madras, Chennai 600036, India\\
}

\affil[2] {Center for Atomistic Modeling and Materials Design, Indian Institute of Technology Madras, Chennai 600036, India \\
}

\author[3]{Srikanth Sastry}
\affil[3]{Theoretical Sciences Unit and School of Advanced Materials, Jawaharlal Nehru Centre for Advanced Scientific Research, Rachenahalli Lake Road, Bengaluru-560064, India\\
}%

\author[4]{Sanat K. Kumar}

\affil[4]{Department of Chemical Engineering, Columbia University, New York, USA\\
}%
\author[1,2]{Tarak K. Patra}

\begin{document}

\maketitle

\begin{abstract}
Polymer melts with chains undergoing reversible crosslinking have distinctively favorable dynamic properties, e.g., self-healing and reprocessability. In these situations there are two relevant elementary time scales: the segmental and the sticker association times. A convenient framework to model these situations is the sticky Rouse model and here we perform hybrid moleculear dynamics (MD) - Monte Carlo (MC) simulations to examine its relevance. In agreement with the underpinning idea discussed above we find that reversibly crosslinked chains show two distinct modes of relaxation behavior depending on the magnitude of bond lifetimes. For bond lifetimes shorter than the chain end-to-end relaxation time, the polymers exhibit essentially Rouse-like dynamics, but with an apparently increased local friction relative to the non-sticky analog. For longer bond lifetimes, the chains exhibit two modes of relaxation: the faster mode is independent of bond lifetime, but the slower mode is controlled by it. However, these slower mode results are not consistent with the predictions of the sticky Rouse model. Our Rouse mode analysis as a function of chain length, $N$, imply that this is likely a result of the relatively short $N$'s employed, but they nevertheless suggest that theories need to include these small chain effects if they are to be relevant to experimental systems with short chains following Rouse dynamics.
\end{abstract}

\newpage

\doublespacing

\section{Introduction}
Permanently crosslinked polymers are ubiquitous due to their favorable mechanical properties. However, they are unsustainable since they cannot be reprocessed. In contrast, polymer networks that can rearrange their topology without depolymerization, e.g., via the formation of reversible crosslinks or by having exchangeable crosslinks (“vitrimers”), are readily reprocessable especially at high temperature, while maintaining their favorable properties under normal use conditions. Understanding the behavior of these constructs is hence an important research topic \cite{montarnal2011silica,lu2012making,jin2013recent,samanta2021polymers,maaz2021synthesis,long2013modeling,meng2016stress,snyder2018reprocessable,semenov1998thermoreversible,rubinstein1998thermoreversible,porath2022vitrimers,zou2018rehealable,arbe2023microscopic,lewis2023effects}. In recent studies, we found that reversible crosslinks can be used to compatibilize immiscible polymer blends\cite{clarke2023dynamic}. Additionally, the saturated liquid density of a reversibly crosslinked polymer increases monotonically with the crosslink fraction, and as a consequence, surface tension increases\cite{karmakar2025computer}. Here, we ask how the relaxation of chains is impacted by the presence of these reversible crosslinks.    
The Rouse model is the simplest description of short chain dynamics in a $\theta$-solvent or in the melt. In this model it is known that the relaxation of a chain can be described as the superposition of normal Fourier modes corresponding to progressively smaller subchain fragments. 
The question of how such chain dynamics are affected by the dynamic exchange of crosslinks is then the question of relevance. 

Work over the last fifty years, beginning with pioneering analytical work of Lodge and coworkers, suggests that the motion of a chain is controlled by the lifetime and concentration of crosslinks.\cite{lodge1956network} For relatively short lived crosslinks (i.e., for bond lifetimes shorter than the relaxation time for chain fragments between two crosslinks), these transient connections serve to simply increase the local friction experienced by the chain beads, thus leading to a predictable increase in chain relaxation time\cite{chen2013ionomer}. In the sticky Rouse model, for longer-lived bonds the chain fragments between crosslinks relax first following Rouse dynamics, and then there is a second mode corresponding to a Rouse relaxation of a chain of stickers with an elementary time scale equal to the bond lifetime. The extension of these ideas to longer entangled chains was then performed by Leibler et al.\cite{leibler1991dynamics}. Another well developed concept is that of an inhomogeneous Rouse model,\cite {stockmayer1975viscoelastic,hansen1975viscoelastic,wang1975dynamics} where monomers have a distribution of frictions. Evidently, the resulting average chain friction is lower than that expected for the standard Rouse by a factor that is determined by the width of the friction distribution\cite{hung2018heterogeneous}.
These ideas were applied to the problem of vitrimers by Ricarte and Shanbag \cite{ricarte2021unentangled} who showed that the local frictions of the normal monomers and sticky monomers added but in an inverse manner in the Rouse description of chain dynamics, along the ideas embodied in the inhomogeneous Rouse model. 

There are several past simulations that have explored these issues \cite{perego2022microscopic,wu2019dynamics,amin2016dynamics,morozova2019properties,zheng2022competing,zheng2023influence,nie2023competing,zhao2024role, perego2021effect, zhao2022molecular, zhao2023unveiling}, and in combination these works show the existence of three distinct relaxations - a bond level event which is controlled by the kinetics of bond formation/destruction/exchange, and two other chain level processes. The faster one of these appears to converge, at long lifetimes, to the relaxation of a permanently crosslinked network, while an even slower mode is evidently driven by slow bond exchange kinetics. Perego and Khabaz examined the volumetric and rheological properties of this class of materials \cite{perego2021effect,perego2022microscopic} and showed, for example, that the (linear) rheology of these materials is controlled by the lifetimes of the transient bonds.
 Xia and de la Cruz \cite{xia2024effect} emphasize the importance of network defects, in particular of dangling ends, in these dynamics.  Wu et al. \cite{wu2019dynamics} examined the dynamics of end-functionalized chains, which are able to reversibly bond to a polyvalent crosslinker. Again, inspite of the relatively short lengths of these chains, these workers found two time scales, one related to the bond lifetime, and the other to the relaxation of larger objects.

With these ideas in mind here we study chain relaxation dynamics of a coarse grained model polymer melt with exchangeable crosslinks using a hybrid molecular simulation method. The bond exchange is simulated using a Monte Carlo (MC) scheme, while molecular dynamics (MD) simulations are conducted to relax the system. An important difference of our work relative to others \cite{perego2021effect,perego2022microscopic,wu2019dynamics,amin2016dynamics} is that bond formation and destruction are independent events in our simulations - in contrast, they are coupled so that the bond exchange between a pair of particles is the dominant (or only permitted) microscopic mechanism in the previous works.
We first characterize chain statistics and the distributions of the distance between two adjacent stickers. The distribution of sticker lifetimes follows an exponential dependence, while previous works which allowed only for bond exchanges, found a stretched exponential relaxation. Subsequently, we study chain dynamics and different modes of relaxations. We consider an uncrosslinked polymer melt and a permanently crosslinked melt as key reference points. The calculations on the corresponding reversibly crosslinked chains show that they are substantially influenced by the lifetime of the reversible bonds. These results closely track previous results for the relaxation of short chains which implies that the details of bond-level events do not qualitatively affect the results obtained \cite{wu2019dynamics}. We analyze the simulation results using Rouse modes, and observe that the chains show two distinctly different types of relaxations when the average lifetime of exchangeable bonds is larger than the longest relaxation time of the corresponding permanently crosslinked system. This last finding is in line with the expectations of the sticky Rouse model, and we examine the role of this model in explaining these findings.

\section{Model and Simulation Protocol}
We use the Kremer-Grest bead-spring polymer model\cite{kremer1990molecular} and perform hybrid MD – MC simulations in an isothermal ensemble as discussed in our previous work\cite{karmakar2025computer}, which adapted earlier ideas of Qin and 
 coworkers\cite{li2022distribution}. The Kremer-Grest coarse-grained model has a venerable history in the polymer community. It balances the fidelity needed to describe chain motion, while at the same time sacrificing chemical fidelity to permit the necessary computational speed-ups needed to study the underlying phenomena. While the resulting information is thus only valid at the Kuhn length scale and larger, these larger scales are precisely where models such as the Rouse or reptation theories are valid. The interaction potential between a pair of non bonded monomers is described by the Lennard-Jones (LJ) 12-6  potential $V(r)=4\epsilon[(\frac{\sigma}{r})^{12}-(\frac{\sigma}{r})^{6}]$ truncated at $r_c$= 2.5 $\sigma$ ($\sigma$ is the monomer diameter and $\epsilon$ is the LJ energy scale). Two successive segments in a chain are connected by a standard finite extensible nonlinear elastic (FENE) bond with $V_{bond}=-0.5kR_{0}^{2}\ln[1-(\frac{r}{R_{0}})^{2}]+4\epsilon[(\frac{\sigma}{r})^{12}-(\frac{\sigma}{r})^{6}]+\epsilon$ where the 2nd term in $V_{bond}$ is truncated at a cutoff distance $r_c = 2^{1/6}\sigma$, with $R_{0}=1.5\sigma$ and $K=30\epsilon/\sigma^{2}$. The total number of monomers in the simulation box is 10000. The chain length is varied from N=10 to 50 in a series of simulations - since the entanglement length for this model has previously been shown to be $N_e \approx$ 70, these chains are effectively unentangled. In this work we specifically focus on two different chain lengths, 20 and 50, respectively. We specifically focus here on the simpler case of unentangled chains because we want to understand the effect of monomer-scale sticker interactions on chain dynamics. Longer entangled chains have an additional length scale corresponding to the tube diameter. We believe that it is prudent to first understand the simpler situation with two length scales before pursuing an understanding of entangled melts.

 \begin{figure}[!htb]
	\centering
	\includegraphics[width=16.3cm,height=8cm]{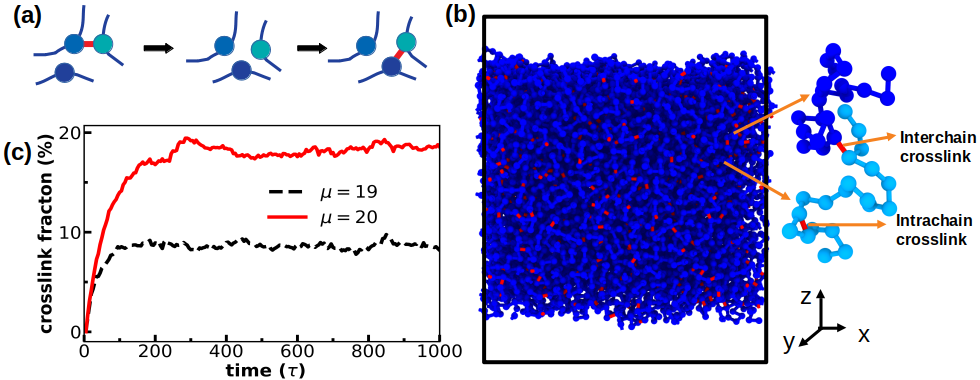}
	\caption{(a) A schematic representation of the dissociate bond exchange mechanism in a polymer network. The dynamic bond is represented as a red line segment between a pair of chains (b) an MD snapshot of the simulation box for chains of length $N$=20. An interchain and an intrachain crosslink are highlighted at the right side of the snapshot. The fraction of particles that are
dynamically crosslinked during a MD is shown in (c) for N = 50 for $\mu=19 \ and \ 20$.}	
	\label{graph0 }
\end{figure}
 We set up our simulation box such that it is periodic in all three directions, but the 3rd direction (z-axis) is significantly longer than the other two so that we have a slab of liquid exposed to vacuum along this direction [Fig. \ref{graph0 }(b)] as in our previous study\cite{karmakar2025computer}. We perform MC cycles at a regular interval during the MD run. The MD time between two consecutive MC cycles is defined as $\tau_{c}$. During each MC cycle, we perform 200 MC moves. Either two randomly chosen monomers within a cutoff distance (1.5$\sigma$) can potentially be connected by a FENE bond as discussed above, or a randomly chosen existing crosslink is selected for deletion during MC moves. The main chain bonds are never created or destroyed. Also, a monomer can participate at most in one crosslink. The bond creation/destruction is accepted following the Metropolis algorithm, where $\mu$ is the chemical potential that determines the equilibrium fraction of crosslinks. The monomers are only allowed to move during the MD, but not during this bond formation/destruction steps. A schematic representation of the dissociative bond exchange mechanism is shown in the Fig. \ref{graph0 }(a). We equilibrate the system first by performing $10^7$ MD steps with an integration timestep of $0.005\tau$, followed by a production cycle of \(3 \times 10^7\) MD steps ($15\times 10^{4}\tau$). Here, $\tau=\sqrt{(m\sigma^{2}/\epsilon)}$ is the unit of time, m, and $\epsilon$ are the mass, size and interaction energy of a monomer, respectively. The temperature of the system $\frac{k_{B}T}{\epsilon}=1$ is maintained by the Nose-Hoover thermostat. MD simulations are conducted using the LAMMPS open-source code.\cite{LAMMPS} Importantly, we note that while our systems correspond to free floating films, we have also conducted simulations on bulk systems without the air-polymer interfaces. While we do find quantitative differences between the two situations especially in their relative behavior, qualitatively, the data behave identically. We therefore proceed with a discussion of the films so as to be consistent with our previously published results\cite{karmakar2025computer}, with this important caveat.

\section{Results and Discussion}
\subsection{Static Properties and Bond Lifetimes}
 The crosslink concentration is a function of the chemical potential. Our simulation protocol generates a polymer network with a fluctuating number of crosslinks, but with an average value that is related to the specified value of the chemical potential $\mu$ [Fig. \ref{graph0 }(c)]. We perform simulations for three chemical potentials, viz.,  $\mu=19,20 $ and $ 22$ (here the chemical potential is made dimensionless relative to the thermal energy). The corresponding crosslink fractions, defined as the average fraction of monomers that are connected by crosslink bonds, are 0.08, 0.18 and 0.46, respectively.  The polymer beads that are part of crosslink bonds are labelled as (transient) sticker beads.  We first measure the distributions of stickers in a chain and their lifetime. To this end, we track sticker beads in an array as a function of $\tau_{c}$. If a bead does not participate in a crosslink we set the value corresponding to it to 0 in an array; otherwise we enter the bonding partner number of that bead. We then calculate the occurrence of a particular number (other than 0) in the array to measure the lifetime of that bond.
\begin{figure}[!htb]
	\centering
	\includegraphics[width=15.3cm,height=12.0cm]{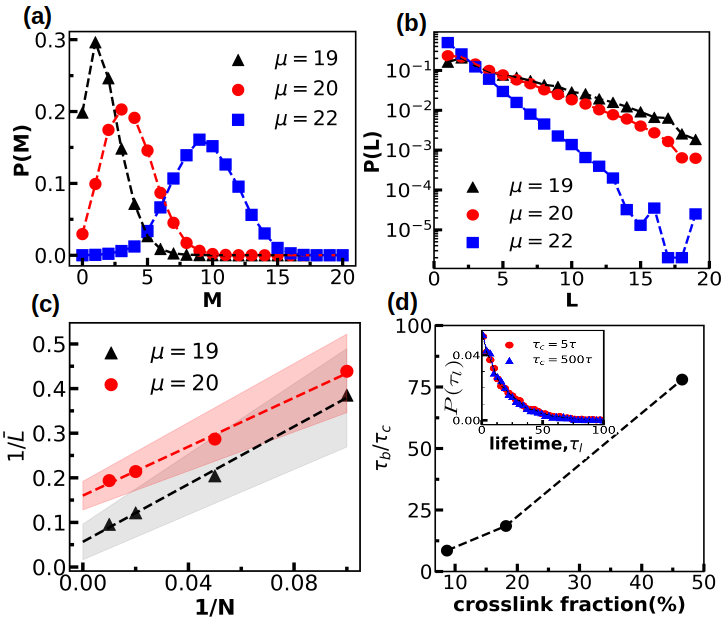}
	
	\caption{ The distributions of the number of stickers in a chain of length $N$=20 is shown in (a) for different chemical potentials. (b) Distributions of the sticker-to-stiker length of a polymer $P(L)$ is shown in (b) for crosslink fractions 8.68\% (black), 18.2\% (red), 46\% (blue) for the chain length $N=20$. (c) Slope $\lambda$ of log$P(L)$ vs $L$ plotted as a function of $1/N$ for $\mu$=19 (black) and 20 (red), respectively. Dashed lines are  linear fits. (d) $\tau_{b}/\tau_{c}$ versus crosslink fraction for $N=20$ with $\tau_{c}=5\tau$. The $\tau_{b}$ is the mean bond lifetime calculated from the inset distribution. Inset: The $\tau_{c}$-normalized bond lifetime distribution $P(\tau_{l})$ for $\tau_{c}=5\tau$(red circle) and $\tau_{c}=500\tau$(blue triangle).}	
	\label{graph1 }
\end{figure}

We quantify the number of stickers in a polymer chain as $M$. The distributions of $M$ for three different $\mu$=19, 20, 22, respectively, in this case of $N$=20, are shown in Figure \ref{graph1 }(a). The peak of the distribution shifts to larger $M$ with increasing $\mu$, as expected. For $\mu=20$, we observe an equilibrium crosslink fraction 0.182 (18.2\%), implying that there are $N(=20)\times0.182 = 3.6$ stickers per chain for the most investigated system. It is to be noted for $\mu$= 19 and 20, 20\% and 3\% chains remain free (with no stickers) on average, respectively. There are no measurable free chains for $\mu=22$.

In addition, we measure the distance ($L$) between two adjacent stickers in a polymer chain (sticker-to-sticker distance) at equilibrium - i.e., the contour length along the chain backbone.  The distribution of $P(L)$ for different $\mu$ is shown in Figure 1b again for $N$=20. The slope $\lambda$ of this distribution is connected to the average $L$ ($\bar{L}$) for an exponential distribution, $\bar{ L}  = 1/\lambda$. We observe that the slope increases with increasing crosslink fraction, as expected. We observe a strong finite chain length effect on $\frac{1}{\bar{L}}$. We explore this further by plotting $\frac{1}{\bar{L}}$ as a function of $1/N$ for two different crosslink fractions, 8.68\%, and 18.2\%, respectively, in Figure \ref{graph1 }(c). For both the cases, we observe a linear dependence from which we can extract the average distance between stickers in the infinite chain length limit from the intercept values. We find intercept values of 0.074 and 0.16, respectively. These numbers are close (with simulation uncertainties) with the determined crosslink fractions of 8.68\% and 18.2\%, respectively. (We also plot a confidence band for the observations which indicates an error bar for the observations.) These results, of course, illustrate the strong chain length dependence of the mean distance between crosslinks. For example, for a crosslink fraction 18.2\% for a chain length $N=20$ with average 3.6 stickers we expect $\bar{L}=5.5$ whereas, we observe $\bar{L}=3.45$ much less than expected. The same trend is observed for the crosslink fraction of 8.68\%. This is likely because the first and last monomers of a chain are not always crosslinked, and hence there are dangling ends which do not contribute to these statistics. 

Next we focus on the mean crosslink bond lifetime. We vary $\tau_{c}$ systematically for a range of $\tau_{c}$=$5\tau$-$1000\tau$ and compute $\tau_{l}$, which is defined as the ratio of bond lifetime and $\tau_{c}$. The distribution of $\tau_{l}$ is shown in the inset of Figure \ref{graph1 }(d) for $\mu=20$. We observe that the $\tau_{l}$ follows an exponential distribution for all $\tau_{c}$. They are also identical for all $\tau_{c}$. We calculate the mean $\tau_{l}$ from this distribution, which is found to be 18.5. It suggests that the mean bond lifetime $\tau_{b}\approx$ 18.5 $\tau_{c}$ for all cases. We plot $\tau_{b}/\tau_{c}$ with increasing $\mu$ or crosslink fraction in Figure \ref{graph1 }(d) for $\tau_{c}=5\tau$. It increases with increasing crosslink fraction, as expected.   \\

\begin{figure}[!htb]
	\centering
	\includegraphics[width=16.3cm,height=8.3cm]{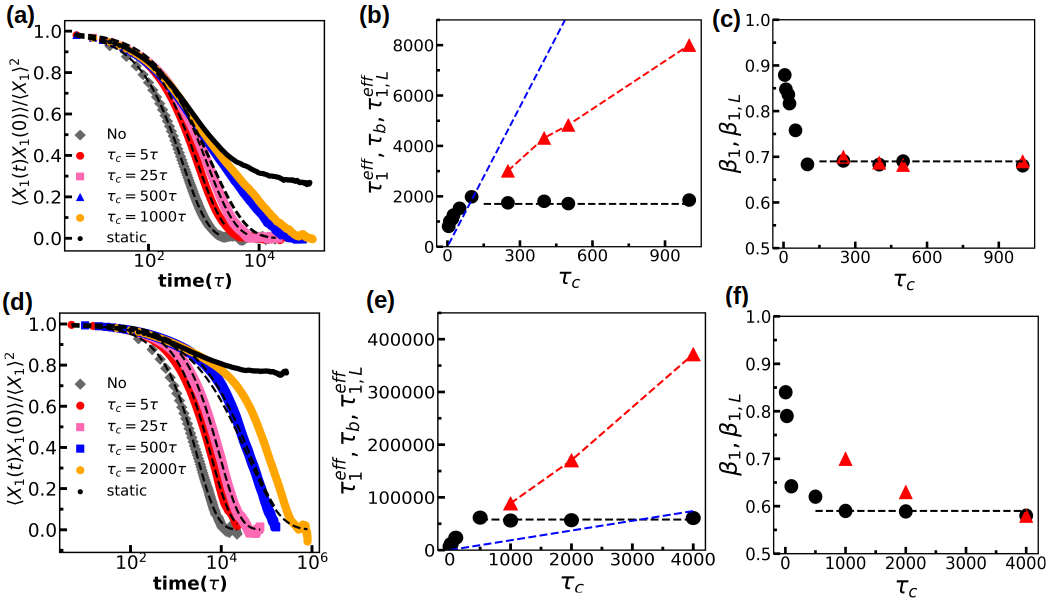}
	\caption{(a)-(c) refer to chains with $N$=20, while (d)-(f) are for $N$=50. Autocorrelation functions of the Rouse mode $p=1$ for different $\tau_{c}$ along with no crosslink and static crosslink systems are shown in (a). The effective relaxation time of the faster process $\tau^{eff}_{1}$ (black),slower process $\tau^{eff}_{1,L}$(red) and average bond lifetime $\tau_{b}$(blue) are plotted as a function of $\tau_{c}$ for crosslink fraction 0.18 in (b). The black dashed straight line around 1800 corresponds to the fast relaxation time of static case for crosslink fraction 0.18. The stretched exponent(faster) $\beta_{1}$  of the faster process and slower process $\beta_{1,L}$ is plotted as a function of $\tau_{c}$ in (c). The black dashed straight line around 0.69 corresponds to the strecth exponent value of fast relaxation of the static case for the crosslink fraction 0.182.}	
	\label{graph3 }
\end{figure}

\subsection{Chain Relaxation and Rouse Modes}
We now consider the relaxation behavior of the chains. We calculate Rouse modes ($p
= 1, 2, ..., N − 1$) of a chain of length N following $\Vec{X}_{p}=(2/N)^{1/2}\sum_{i=1}^{N} \Vec{r}_{i} cos[(p\pi/N)(i-1/2)]$\cite{kalathi2014rouse}. The modes describe internal relaxations with a mode number $p$ corresponding to a chain fragment that consists of $(N − 1)/p$ bonds, where $p=1$ corresponds to the longest mode, also called end-to-end relaxation. The autocorrelation of the Rouse modes, termed as Rouse relaxations, is observed to follow a stretched exponential temporal dependence of the form $\langle \Vec{X}_{p}(t)\cdot \Vec{X}_{p}(0) \rangle = \langle \Vec{X}_{p}^{2} \rangle e^{-(\frac{t}{\tau_{p}})^{\beta_{p}}} $ in past works. Here, $\tau_{p}$ and $\beta_{p}$ are the Rouse relaxation time and the exponent of mode $p$, respectively. We also compute the effective relaxation time for each mode, defined as $\tau_{p}^{eff} = \frac{\tau_{p}}{\beta_{p}}\Gamma(1/\beta_{p})$. To this end, we first calculate the end-to-end distance relaxation function, i.e., the mode corresponding to $p$=1 in the Rouse model for the uncrosslinked chains, and the permanently crosslinked case (for a crosslink fraction of 18.2\%), respectively, Figure \ref{graph3 }a. While the melt system relaxes fully, the permanently crosslinked system shows relaxation to a plateau that does not further decay (as expected). Note that the plateau value is higher for the $N$=50 system relative to the $N$=20 case. Since the Flory prediction for the critical gelation threshold is one crosslink per chain, or an average of 2 monomers per chain that participate in crosslinks, $N$=20 case with $\approx$ 3.6 crosslinked monomers per chain is well past its gel point. Similarly, for $N$=50 we expected $\approx$ 9.1 crosslinks per chain. The inability for the permanently crosslinked system to relax fully thus refers to the single percolating graph which is fixed, while the relaxations that do occur correspond to the dangling ends in this network. 

In this context, it is thus interesting that the end-to-end relaxation functions fully relax for all the dynamically crosslinked systems studied (which have the same crosslink density as the permanent case). Clearly, all our simulations are long enough to relax the chains and achieve equilibration (Figure \ref{graph3 }a). We fit this correlation function to a single or double stretched exponential function of the form $\frac{\langle \Vec{X}_{p}(t)\cdot \Vec{X}_{p}(0) \rangle}{\langle \Vec{X}_{p}^{2} \rangle} = A e^{-(\frac{t}{\tau_{1}})^{\beta_{1}}} $. This model is inspired by the ideas of Ricarte and Shanbag \cite{ricarte2021unentangled} who predicted that the correlation function should be a sum of two exponential relaxations  - one for the normal Rouse chain (between stickers) and the longer sticky chain relaxation time scale. The only difference here is that we employ two stretched exponential processes to describe these relaxations.

We observe rich behavior in this relaxation process over different ranges of the crosslink lifetimes - in particular, we systematically vary the attempt times, $\tau_{c}$. First, for $N$=20 for $\tau_c \le 150\tau$, the data can be fit by a single relaxation process, while for larger $\tau_c$ two relaxations are evident.  A similar result is found for $N$=50, with the crossover happening for $\tau_c\approx 500\tau$. This result is surprising since this crossover, in the framework of the sticky Rouse model, should happen when the relaxation time for the chains becomes smaller than the Rouse relaxation time of a chain of stickers. Thus we expect that $N_s^2\tau_b\sim N^2 \tau$ at the crossover. Here $N_s$ is the number of stickers per chain, and hence $N/N_s=1/0.182\approx 5.5$ is independent of chain length, implying that the crossover should happen at roughly the same value of $\tau_b\approx 30\tau$. These results likely arise because the monomer relaxation time $\tau$ depends on the presence of the stickers. This is seen clearly in Figure \ref{graph3 }b and \ref{graph3 }d where the relaxation time of the chains, $\tau_{1}^{eff}$, increases essentially linearly with $\tau_c$ at small $\tau_c$, implying that the sticker interactions act as additional monomer friction in this limit. 
A second fact, embodied in Figure \ref{graph1 } is that the length of the chains between stickers is smaller than expected because of finite chain length effects. Since this quantity increases with increasing $N$, this might also help to explain these findings.

Evidently, $\tau_{1}^{eff}$ increases and saturates for $N$=20 around $1800\tau$ for $\tau_{c}\ge 150\tau$, and for $N$=50 around $8000\tau$ for $\tau_c\ge 3000\tau$.
This saturation value closely matches the relaxation time of the permanently crosslinked analog (which has a long time plateau in the correlation function and hence does not relax fully). These results are reminiscent of the recent work of Wu et al. \cite{wu2019dynamics} even though their model is quite different than ours. 

A slower relaxation starts to emerge around $\tau_{c}=150\tau$ - the resulting  timescale, $\tau_{1,L}^{eff}$, increases monotonically (and approximately linearly) with $\tau_c$, Figure \ref{graph3 }(b). 
Clearly, the relaxation process splits into a slow and a fast process when the bond lifetime becomes longer than the single measurable relaxation time of the permanently crosslinked system. We suspect for larger $\tau_{c}>150 \tau$, the fast process comes due to the relaxation of the dangling ends of the network, while the longer process, which is almost a linear extrapolation of short $\tau_{c}$ relaxation, is likely related to the sticky Rouse dynamics of the chains.

A surprising finding here is that the slower Rouse mode for $N$=20 is considerably faster than the arithmetic mean lifetime of a single sticky bond, which is shown as a dotted blue line in Figure \ref{graph3 }(b). Thus, the longest relaxation times do not appear to follow the sticky Rouse model prediction, $\tau_{b} N_{s}^{2}$, where $\tau_{b}$ is the average bond lifetime and $N_{s}$ is the average number of stickers per chain. This reverses for $N$=50 and the slower process is considerably slower than the bond lifetime, while still being about 5 times smaller than the sticky Rouse prediction. We conclude that these are finite chain length effects, and these corrections need to be included in any new theory of chains with sticky interactions.

We also plot in Figure \ref{graph3 }(c) the stretching exponents for the two processes, $\beta_{1}$(faster process) and $\beta_{1,L}$(slower process). $\beta_{1}$ starts from a high value, $\approx$ 0.86 for small $\tau_{c}$ and decreases and saturates around 0.7 beyond $\tau_{c}=100\tau$. We observe  $\beta_{1}$ at large $\tau_{c}\ge150\tau$ matches with the relaxation of the permanently crosslinked system.  We also plot the stretching exponent of the slower process $\beta_{1,L}$ as a function of  $\tau_{c}$ in Figure \ref{graph3 }(c). We observe $\beta_{1,L}$ remains almost constant at $\approx 0.7$ and approximately equals the value of $\beta_{1}$ for the faster relaxation.

\begin{figure}[!htb]
    \centering
     \includegraphics[width=16.3cm,height=8.3cm]{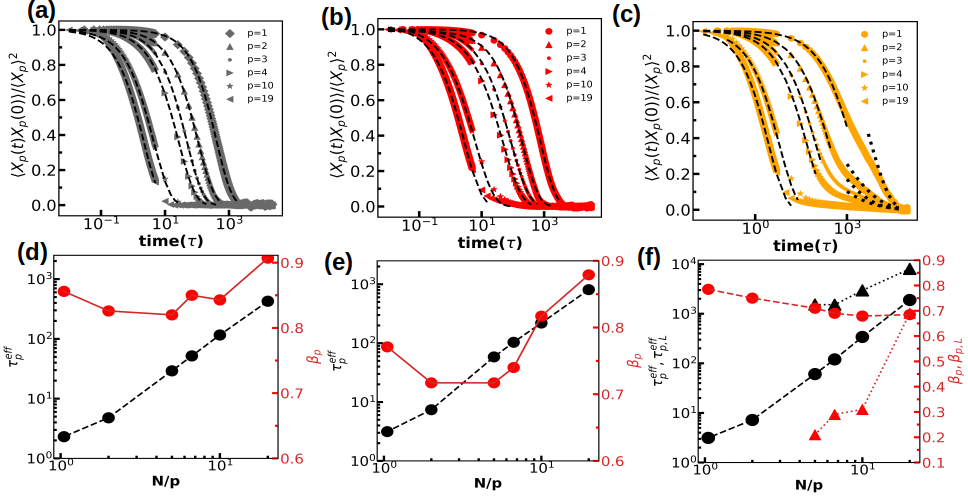}
    \caption{Normalized autocorrelation function of different Rouse modes $p$ for no crosslinked system, dynamic crosslinked system with  $\tau_{c}=5\tau$, and $\tau_{c}=1000\tau$ are shown in (a), (b), and (c), respectively. Effective relaxation times of all the modes $\tau^{eff}_{p}$ for no-crosslined, crosslinked system with  $\tau_{c}=5\tau$, and $\tau_{c}=1000\tau$ are shown in (d), (e), and (f), respectively. The fraction of crosslinked monomer is 18.2\%, and the chain length is $N=20$ for all the cases. The right y-axes of (d), (e) and (f) correspond to the $\beta_{p}$ values of the respective fitted stretched exponent curves. In panel (f), the effective relaxation times $\tau^{eff}_{p,L}$, $\beta_{p,L}$ with $N/p$ for the slower process for crosslink fraction 18.2\% with $\tau_{c}=1000\tau$,$N=20$ are also plotted.}
    \label{graph4 }
\end{figure}

Next, we explore the different Rouse modes for $N=20$, especially their effective relaxation times, $\tau^{eff}_{p}$, and their stretching exponents, $\beta_{p}$, for a system with 18.2\% dynamic crosslinks with $\tau_{c}=5\tau$ and $\tau_{c}=1000\tau$, respectively. These are compared to the corresponding melt with no crosslinks. In the neat uncrosslinked melt, $\tau^{eff}_{p}$ follows the Rouse scaling with $N/p$, Figure \ref{graph4 }(a) and (d). In more detail, the $\tau^{eff}_{p} p^2$ decreases slightly with increasing $N/p$, so that the Rouse scaling is only approximately followed. 
This matches with findings in the literature. The stretching exponents $\beta_{p}$ vary with  $N/p$, with a strong departure from the Rouse result of $\beta_p$=0.5. A previous work\cite{kalathi2014rouse} by us has shown that $\beta_p$ results for small values of $N/p$ are indeed closer to $\approx$ 0.75-0.8, with values of $\approx$ 0.5-0.6 only emerging for long chains where we can access modes with $N/p \ge \approx$ 70. Clearly, our results are a manifestation of the short chains employed in these calculations. 

Now, for the dynamically crosslinked system with $\tau_{c}=5\tau$, we observe similar trends as in the uncrosslinked case for $\tau_{p}$ and $\beta_{p}$, Figure \ref{graph4 }(b) and (e). The slope of $\tau_{p}^{eff}$ $vs.$ $N/p$ is still consistent with the Rouse scaling, but the absolute values of the relaxation times are higher. Effectively, at this $\tau_c$, the reversible crosslinks act like they increase the effective monomer friction coefficient without qualitatively altering the mode relaxations of the chains. Correspondingly, the $\beta_{p}$ goes from 0.7 to 0.86 with changes in $N/p$. 
The system with $\tau_c$=1000$\tau$, on the other hand, shows a two steps relaxation for $p=1$ and also for higher modes $p$=2,3 and 4, Figure \ref{graph4 }(c) and (f). For even higher modes $p=10,19$, the relaxation goes back to being monomodal. For the slower process, both the effective time $\tau_{p,L}^{eff}$ and stretching exponent $\beta_{p,L}$ decrease with decreasing $N/p$.

\begin{figure}[!htb]
	\centering
	\includegraphics[width=16.3cm,height=5.0cm]{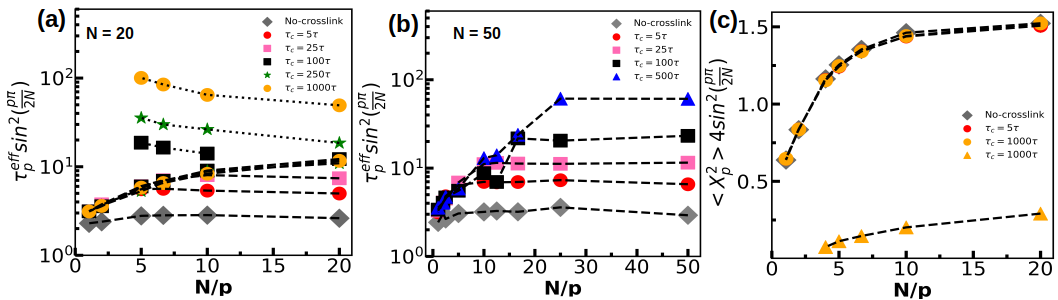}
	\caption{ The $\tau_{p}^{eff}sin^{2}(p\pi/2N)$ is plotted as a function of $N/p$ for chain length $N=20$ for different bond lifetime of crosslink fraction 18.2\% and for no crosslink case in (a). The $\tau_{p}^{eff}sin^{2}(p\pi/2N)$ is plotted as a function of $N/p$ for the chain length $N=50$ for different bond lifetime of crosslink fraction 18.2\% and for no crosslink case in (b). The scaled Rouse amplitude $\langle X_{p}^{2} \rangle 4 sin^{2}(p\pi/2N)$ for different modes over $N/p$ is shown in (c) for the chain length $N=20$. For $\tau_{c}=1000\tau$, square symbols represent the amplitude for the faster process and triangle symbols correspond to the slower process.}
	\label{graph5 }
\end{figure}

First, we comment on the fast relaxation time as a function of $N/p$ and $\tau_c$. Clearly, at a given $N/p$, the effective friction $\zeta$, which is defined as $\frac{\zeta b^2}{12 k_BT} = \tau_p sin^2\left ( \frac {p \pi}{2N} \right)$,
increases with increasing $\tau_c$, Figure \ref{graph5 }a. Here $b$ is the Kuhn length of the chains. This is as expected, and is a consequence of sticker interactions acting as increased friction. The effective friction increases with $N/p$ at a given $\tau_c$, Figure \ref{graph5 }a. While the uncrosslinked melt results show a plateau at large $N/p$, and this trend is also follows for small $\tau_c$, no plateau is found for the two largest values of $\tau_c$. When compared to the uncrosslinked melt, these results clearly show that the effective friction becomes larger when larger $N/p$ values are considered. One potential means to understand this result is that more stickers are present in longer chain segments, i.e., as we consider larger $N/p$. In contrast to these results, the scaled amplitude of the Rouse modes, ($\langle  X_p^2 \rangle   4 sin^{2}(p\pi/2N)$) do not show any dependence on $\tau_c$, Figure \ref{graph5 }c. These amplitudes do increase with increasing $N/p$, but this is as expected from past works. For $\tau_{c}=1000\tau$ we also show amplitudes for slower process.
We also identify strong finite chain length effects on these fast relaxations. We calculate the effective frictions for $N=50$ and find the curves for smaller $\tau_{c}$ follows similar trends, but that we do see plateau values for $\tau_c=500 \tau$, the largest values of $\tau_c$ that we have studied, Figure \ref{graph5 } b. We now turn to the slower relaxation modes. The effective friction is higher for the slower mode than for the fast mode, except they seem to converge for smaller $\tau_c$ values at large $N/p$. Going beyond this, the effective friction increases with decreasing $N/p$ (or the Rouse rate decreasing with decreasing $N/p$). 

\subsection{Discussion}
We first comment on the distribution of bond lifetimes. In our work, we find that the distribution of bond lifetimes follow a simple exponential dependence. This finding should be contrasted with past works which have found a stretched exponential relaxation function \cite{perego2021effect,perego2022microscopic,wu2019dynamics}. We note that these past works impose bond swapping as an essential mode of relaxation at this scale. Our work, on the other hand, allows bond formation and breaking to occur as independent events. This essential difference might be at the root of the different dependencies (stretched vs normal) that the distribution of bond lifetimes follow. Regardless of these differences in microscopic behavior, all of these works find that the fast relaxation behavior seen in Figure \ref{graph3 }b, i.e., a saturation of this time to the value expected for a permanently crosslinked network for long enough bond lifetimes. We thus conclude that the details of the microscopic bond exchange processes do not affect chain level processes. This is an important finding which reveals the universal aspects of this relaxation behavior. Finally, we comment on the slow process that is seen in our work. We note that while Amin et al.\cite{amin2016dynamics} do observe a slow process, the previous works by Wu et al. \cite{wu2019dynamics} and Khabaz \cite{perego2021effect,perego2022microscopic} do not observe this slower process. We attribute this to the fact that Khabaz et al do not run their simulations long enough to observe this slower process. It remains unclear to us as to why Wu et al. do not observe this slower process which evidently accompanies the formation of a transient network with long lifetimes.

\section{Conclusions} 
Reversibly crosslinked polymers  have the unique ability to combine high mechanical properties with reprocessibility. Here, we report the relaxation behavior of such reversibly crosslinked systems through coarse-grained molecular dynamics simulations. Our results point to the importance of bond lifetime in the chain relaxation. We observe that there is a rich difference in chain dynamics for long lived bonds compared to short lived ones. In the latter case, we recover Rouse dynamics, but with an increased friction. For sufficiently long lived bonds, we find a second slower relaxation, but its behavior is not consistent with the expectations of the sticky Rouse model. This is likely due to finite chain length effects, and hence we  emphasize the need for improved models to describe the relaxations of these chains with reversible crosslinks. \

\section*{Conflicts of interest}
There are no conflicts to declare. \

\section*{Acknowledgments}

This work is made possible by financial support from the SERB, DST, Government of India through a core research grant (CRG/2022/006926). This research uses the resources of the Center for Nanoscience Materials, Argonne National Laboratory, which is a DOE Office of Science User Facility supported under the Contract DE-AC02-06CH11357.

\bibliographystyle{vancouver}

\end{document}